\documentclass[useAMS]{mn2e}
\usepackage[]{txfonts}
\def\simless{\mathbin{\lower 3pt\hbox
{$\rlap{\raise 5pt\hbox{$\char'074$}}\mathchar"7218$}}}   
\def\simmore{\mathbin{\lower 3pt\hbox
{$\rlap{\raise 5pt\hbox{$\char'076$}}\mathchar"7218$}}}   
\newcommand{\be}{\begin{equation}}
\newcommand{\ee}{\end{equation}}
\usepackage{graphicx}
\usepackage{epsfig} 
\usepackage{epstopdf}

\title[Flares from Galactic centre pulsars]{
Flares from Galactic centre pulsars: a new class of X-ray transients?
}

\author[Giannios \& Lorimer]
{Dimitrios Giannios$^{1}$\thanks{E-mail: dgiannio@purdue.edu  (DG)}  and Duncan R.~Lorimer$^{2}$\thanks{E-mail: duncan.lorimer@mail.wvu.edu (DL)}  \\
$^{1}$Department of Physics and Astronomy, Purdue University, 525 Northwestern
Avenue, West Lafayette, IN 47907, USA\\
$^{2}$Department of Physics and Astronomy, West Virginia University, Morgantown, WV 26506, USA}

\begin{document}
\date{Received / Accepted}
\pagerange{\pageref{firstpage}--\pageref{lastpage}} \pubyear{2013}

\maketitle

\label{firstpage}

\begin{abstract}
Despite intensive searches, the only pulsar within 0.1~pc of the
central black hole in our Galaxy, Sgr~A*, is a radio-loud
magnetar. Since magnetars are rare among the Galactic neutron star
population, and a large number of massive stars are already known in
this region, the Galactic centre (GC) should harbor a large number of
neutron stars. Population syntheses suggest several thousand neutron
stars may be present in the GC. Many of these could be
highly energetic millisecond pulsars which are also proposed to be
responsible for the GC gamma-ray excess. We propose that the presence
of a neutron star within 0.03~pc from Sgr~A* can be revealed by the
shock interactions with the disk around the central black hole.
As we demonstrate, these interactions result in observable transient non-thermal
X-ray and gamma-ray emission over timescales of months, provided that
the spin down luminosity of the neutron star is $L_{\rm sd}\sim 10^{35}$~erg~s$^{-1}$. 
Current limits on the population of normal and millisecond pulsars
in the GC region suggest that a number of such pulsars are present
with such luminosities.
\end{abstract} 

\section{Introduction} 
\label{intro}

The compact radio source Sgr~A* is believed to be the location of
the massive black hole in the Galactic centre (GC) of mass $M_{{\rm
BH}}=4.3\times 10^{6}$\,M$_{\odot}$ (and corresponding
gravitational radius $R_{\rm g}=GM_{\rm BH}/c^2\simeq 6.4\times
10^{11}$~cm). {\it Chandra} has resolved the hot gas surrounding
Sgr~A* at a scale of $10^{17}$~cm (Baganoff et al.~2001).
Bremsstrahlung emission from this region accounts for much of the
observed quiescent soft ($T\sim 1$~keV) X-rays of luminosity $L_x\sim
2\times10^{33}$~erg~s$^{-1}$. The inferred gas density at this
distance $n\sim 100$~cm$^{-3}$ (Baganoff et al.~2003).  Accretion onto
the black hole is likely to power an IR source of luminosity $\sim
10^{36}$~erg~s$^{-1}$ (Genzel et al.~2010).  Both IR and X-ray flaring
is observed on timescales ranging from minutes to
hours. Flares are believed to be associated with processes that take
place at the inner accretion disk, close to the black-hole horizon 
(Baganoff et al.~2001; Ghez et al.~2004).

Discovering radio pulsars in the GC is one of the holy grails in
astrophysics due to their promise as probes of the central
supermassive black hole (see, e.g., Psaltis et al.~2015), and in
deciphering the nature of the interstellar medium in its vicinity
(Cordes \& Lazio 1997). The inner pc in the GC is expected to have
hundreds of radio emitting pulsars (see, e.g., Pfhal \& Loeb 2004;
Wharton et al.~2012). The discovery of a magnetar within $\sim
0.1$~pc srengthens the case for their presence (Mori et al.~2013;
Eatough et al.~2013). In addition, some 20 massive stars are known to
orbit Sgr A* within $10^{17}$~cm (Ghez et al.~2004). Such stars are
expected to give birth to pulsars. Further evidence for a significant
population of millisecond pulsars (MSPs) in the GC region was recently
presented by Brandt \& Kocsis (2015). These authors demonstrate that
the 2~Gev excess gamma-ray emission from the GC detected by {\it Fermi} 
is consistent with an ensemble of order 1000~MSPs in this
region.  Similar conclusions were also reached by other authors (Qiang
\& Zhang 2014). It should be noted, however, that the {\it Fermi}
results can also be explained by a dark matter model (e.g.,
Hooper \& Goodenough 2011). A significant effort is being invested
into further searches for dark matter in the GC (see, e.g., van Eldick 2015).

One way to discriminate between the MSP and dark matter scenarios would be the
detection of MSPs in the GC. Macquart \& Kanekar (2015)
predict optimal results for future radio surveys in the 8~GHz
band. Although such searches are now being carried out, their success
is strongly dependent on the scattering environment in the GC region
which is still not well understood. The principle focus of this paper
is to develop a new approach to constraining the pulsar population in
the GC.  As shown schematically in Fig.~\ref{fig:1}, we propose that
when a neutron star (NS) approaches within $\sim 0.03$~pc from Sgr~A*,
the ram pressure from the accretion disk shocks the pulsar wind fairly
close to the NS. Pairs accelerated at the termination shock power a
non-thermal, synchrotron flare that will last for months to years
and can be detected in the X-ray band and, possibly, at other wavelengths.

This {\it Letter} is organized as follows. In \S 2 we detail the emission 
processes of our model. In \S 3 we review the evidence for a significant
number of NS in the GC region. In \S 4 we discuss prospects for observing
this population via our proposed mechanism.

\section{Transient X-ray Emission from Galactic Centre Neutron Stars}

\begin{figure}
\includegraphics[scale=0.45]{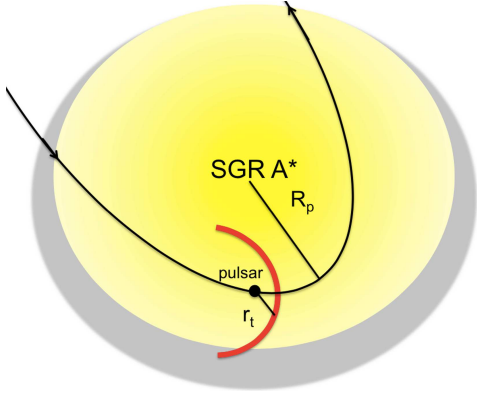}
\caption{Sketch showing the essence of our model.  A NS
  passing through the accretion disk around Sgr~A*, with pericentre
  distance $R_p$, has a wind that is sufficiently energetic to produce a
  synchrotron emitting, bow-shock region with termination distance $r_t$ 
  which can be observable at X-ray energies.}
\label{fig:1}
\end{figure}

NSs residing in the inner 0.1~pc of the GC are characterized by
fast motions with the winds undergoing strong interactions with the
gas surrounding Sgr~A* and its accretion disk.  Here we demonstrate
that when the NS dives into the accretion disk, its relativistic
wind is shocked, giving a powerful nonthermal
X-ray and gamma-ray transient that lasts from $\sim$~1 month to
years. Pulsar winds are known to be bright X-ray and $\gamma$-ray
sources when they interact with an ionized medium (e.g., Romani et
al.~1997). These interactions are particularily intense in systems
where the pulsar has a close, massive star companion (Paredes et
al. 2013; Dubus 2013). The interaction model we propose, shown schematically in
Fig.~\ref{fig:1}, is based on earlier work by Giannios \& Sironi
(2013) can be applied to any type of spin-powered pulsar.

We consider a NS with spin down luminosity $L_{\rm sd}$ in an
eccentric orbit around the central supermassive black hole with
pericentre passage at distance $R_p \ll 1$~pc as shown in
Fig.~\ref{fig:1}. At $R\sim 10^{17}$~cm, the gas density is probed by
its bremsstrahlung emission to be
of order $n\sim 100$~cm$^{-3}$ assuming a modest gas filling
factor of $\sim 1/4$ at this scale (Baganoff et al.~2003).
At $R<10^{17}$cm, the gas may have sufficient angular momentum 
to form an accretion disk. Because of the low accretion rate to the
black hole, the accretion is likely to take place through a 
geometrically thick disk with its density profile being rather model dependent. 
Possible scalings range from $n\propto R^{-3/2}$ as, e.g., motivated
by the advection dominated accretion flow solution (Narayan et
al.~1995) or $n\propto R^{-1}$ as, e.g., motivated by general
relativistic magnetohydrodynamic simulations (Tchekhovskoy \&
McKinney~2012).  In the case of a convection dominated accretion flow
solution, the density profile is shallower ($n\propto R^{-1/2}$;
Quataert \& Gruzinov 2000).  For the estimates that follow we adopt
an intermediate profile $n=1000/R_{16}$~cm$^{-3}$, where
$R=10^{16}R_{16}$~cm. A similar density profile is expected
if one allows for moderate mass-loss in the disk through
winds (Blandford \& Begelman 1999).

Assuming a highly elliptical orbit, 
the characteristic velocity of the neutron star at pericentre
\begin{equation}
v_p \simeq \sqrt{2 R_g/R_p}c \sim 3.4\times10^{8}R_{16}^{-1/2} 
\,{\rm cm~s}^{-1}.
\end{equation}
This expression also holds for 
quasi-circular orbits, within a factor of two.
At pericentre, the ram pressure $\rho v_p^2\simeq 1.9\times 10^{-4}R_{16}^{-2}$~cgs 
of the disk is maximal and shocks the NS
wind at a distance from the NS at which it matches the wind
pressure, $L_{\rm sd}/4\pi r^2c$. The resulting termination distance 
from the NS,
\begin{equation}
r_t = \sqrt{ \frac{L_{\rm sd}}{4 \pi \rho v_p^2 c} }
\sim 3.7 \times 10^{13} L_{35}^{1/2}R_{16}~{\rm cm}.
\end{equation}
The shocked wind moves at a mildly relativistic speed $\sim c/2$ 
and expands on a timescale 
\begin{equation}
t_{\rm exp} \sim 2r_t/c\sim 2.5\times 10^3L_{35}^{1/2}R_{16}~{\rm s}.
\end{equation}

At the termination shock, pairs are expected to be accelerated to provide a
non-thermal particle distribution. We set the particle distribution 
$N(\gamma)\propto \gamma^{-2}$ for $\gamma\ge \gamma_{min}\sim 10^5$ up to a maximum synchrotron
burnoff limit (de Jager et al. 1996).  The magnetic field pressure at the shocked
region can be parameterized as a fraction $\epsilon_B$ of the gas
pressure. {Assuming rough equipartition between 
particles and magnetic field at the shock downstream (see, e.g.,
Porth, Komissarov \& Keppens 2014):}   $B\sim 0.04\epsilon_{B,1/3}^{1/2}R_{16}^{-1}$G.
Electrons accelerated to $\gamma_x\sim 3.5\times
10^6\nu_{5keV}^{1/2}\epsilon_{B,1/3}^{-1/4}R_{16}^{1/2}$ 
by the shock will emit synchrotron emission in the $\nu=5$~keV range and cool on a timescale
\begin{equation}
t_{\rm syn} \simeq \frac{7.5 \times 10^8}{\gamma_xB^2} {\rm s}\simeq 1.3 \times 10^5 R_{16}^{3/2}\epsilon_{B,1/3}^{-3/4}\nu_{\rm 5~keV}^{-1/2}~{\rm s}. 
\end{equation}
The power radiated in the X-ray band is, therefore, 
\begin{equation}
L_x\sim (t_{\rm exp}/t_{\rm syn})L_{\rm sd}\simeq 2\times 10^{33}
L_{35}^{3/2}\epsilon_{B,1/3}^{3/4}R_{16}^{-1/2}\nu_{\rm 5~keV}^{1/2}~
{\rm erg}~{\rm s}^{-1}. 
\end{equation}
In this framework, the emission from a
pulsar of $L_{\rm sd}\sim 10^{35}$~erg~s$^{-1}$ at a pericentre passage of
$R_p \leq 10^{17}$~cm powers detectable X-ray flares
because: (i) their X-ray luminosities are comparable to the quiescent
emission from Sgr~A*; (ii) their spectra are distinctly
non-thermal extending into the hard X-ray band.

Figs.~2 and 3 show the synchrotron emission spectra at pericentre
for different spin down luminosities of the neutron star and
pericentre distances $R_p$. At photon energies below $E_{\rm
 min}\simeq 4\gamma_{\rm min,5}^2 \epsilon_{B,1/3}^{1/2}R_{16}^{-1}$~eV, the emission 
spectrum slope scales as $f_E\propto E^{1/3}$
with the break energy detemined mainly by the minimum Lorentz factor
of the electron distribution behind the termination shock 
$\gamma_{\rm min}$. For $E>E_{\rm min}$, the flux  $f_E\propto E^{-1/2}$ up to
energy $E_{\rm c}\simeq 16 L_{35}^{-1}\epsilon_{B,1/3}^{-3/2}R_{16}$~MeV
above which the particles are cooling rapidly. For $E\simmore E_{\rm c}$, the pairs radiate at a rate
 comparable to the spin-down luminosity of the pulsar
for the adopted particle power-law slope of $-2$. This estimate is likely to be
optimistic for the $\gamma$-ray emission from the transient since, 
for a steeper particle index $p<-2$, the $\gamma$-ray emission is weaker.
The luminosity from the shocked wind at pericentre can 
easily exceed that of the quiescent X-ray emission and, under
favorable conditions, be comparable to that in the {\it Fermi-LAT} band.

The pericentre passage of the pulsar takes place over a timescale
\begin{equation}
t_p\sim R_p/v_p\sim 3\times 10^7 R_{16}^{3/2}~{\rm s}.
\end{equation}
This is also the characteristic timescale over which the pulsar wind 
undergoes the most intense interaction with the surrounding accretion
disk and, as a result, marks the duration of the flares.
Whether one expects a single flare or a pair of flares, depends on
the relative inclination of the orbit of the neutron star with respect to
that of the accretion disk (see Giannios \& Sironi 2013).  If the disk
is co-planar with the stellar orbit (e.g., they are co- or counter-rotating)
the emission from the bow shock will peak at pericentre. For large
inclination of the two orbital planes, two
flares are expected: one before and one after the pericentre passage. 
Since  for thick disks, where the height is comparable to the radius,
the timescale for each crossing of the 
midplane of the disk is similar to the duration of the pericentre
passage $t_p$ (for lightcurve examples, see Giannios \& Sironi 2013).

The previous estimates of the synchrotron emission from the shocked pulsar
wind are based on a simple one-zone model. 
The shocked fluid follows a bow shock structure. 
The accelerated particles radiate substantially 
close to the termination shock but they continue to radiate further back
in the tail. The pulsar spends at pericentre much
longer the time it takes for the shocked fluid to leave the
termination shock: $t_p/t_{\rm exp}\sim 10^4R_{16}^{1/2}L_{35}^{-1/2}$. In the tail of
the bow shock, the shocked fluid is still confined by the thermal
pressure from the accretion disk. The pressure of the disk is
not negligible $P_d\sim 0.5 \rho_d v_k^2\sim P_{\rm ram}/4$;
containing substabtial turbulent magnetic field of strength $B_d\sim 0.02
\epsilon_{B,1/3}^{1/2}R_{16}^{-1}$~G. Therefore,
after a modest initial expansion, the shocked wind will undergo a
much longer process of mixing with the disk material. Substantial
emission for the ultrarelativistic particles is expected at this stage
possibly enhancing the power of the transient. The calculation
of the emission from this region requires a more detailed
hydrodynamical calculation that is beyond the scope of this {\it Letter}.

\begin{figure}
\includegraphics[scale=0.4]{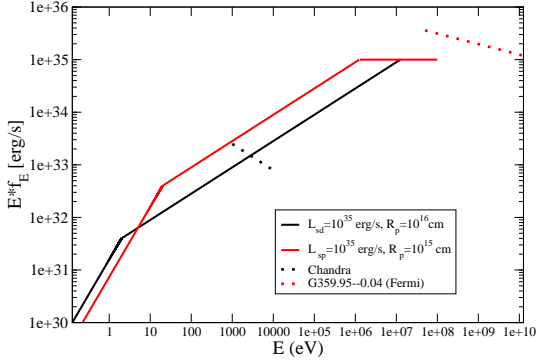}
\caption{Synchrotron spectra from the shocked pulsar
wind for neutron star orbit with pericentre distance 
$R_p=10^{16}$~cm (black curve) and $R_p=10^{15}$~cm (red curve),
respectively. The pair minimum Lorentz factor $\gamma_{\rm min}=10^5$
and the spin down power $L_{\rm sd}=10^{35}$~erg~s$^{-1}$. The maximum
sunchotron energy is set at $\sim 100$~MeV. The black dotted line
shows the Chandra quiescent emission from Sgr A* and the 
red, dash-dotted line the {\it Fermi} level of emission seen for the
source G359.95--0.04 at the same region.}
\label{fig:2}
\end{figure}

\begin{figure}
\includegraphics[scale=0.4]{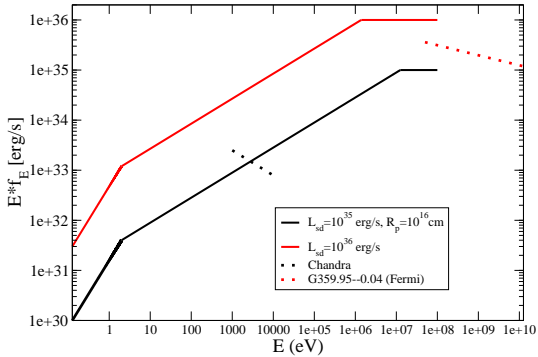}
\caption{Same as Fig.~1 but for pulsar spin down luminosity
$L_{sd}=10^{35}$~erg~s$^{-1}$ (black curve) and
$L_{sd}=10^{36}$~erg~s$^{-1}$, respectively. The NS has
orbit with precentre distance $R_p=10^{16}$~cm.}
\label{fig:3}
\end{figure}

\section{Neutron stars within 0.1 parsec of Sgr~A*}

\begin{figure}
\includegraphics[scale=0.3,angle=270]{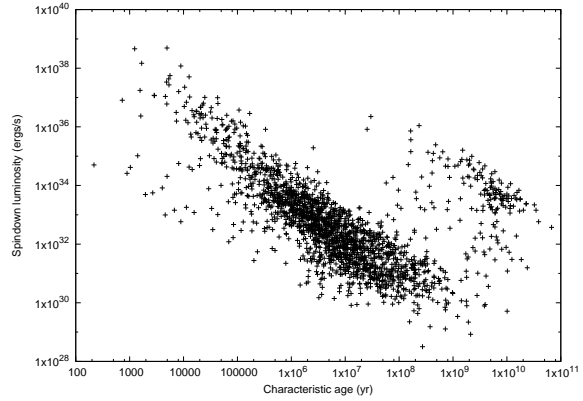}
\caption{Scatter diagram showing spin-down energy loss rate ($\propto
  \dot{P}/P^3$) versus characteristic age ($\propto P/\dot{P}$) for a
  sample of 1982 pulsars take from the ATNF pulsar catalog (Manchester
  et al.~2005).}
\label{fig:edotage}
\end{figure}

{As demonstrated above, pulsars and MSPs with spin down luminosity
$L_{\rm sd}\sim 10^{35}$~erg~s$^{-1}$ and with orbit with pericentre distance
of $r_{\rm p}\sim 0.01$~pc from Sgr~A* make promising sources for an observable
transient. Given the highly elliptical orbits of the stars observed in the
region (see next Section for details),  these NS are expected to
spend most of their orbit at apocentre distance of 
$r_{\rm app}\sim 0.1 \rm pc$.} As can be seen from the current sample of pulsars shown in
Fig.~\ref{fig:edotage}, a significant fraction, approximately 7\% of
all observed pulsars, have $L_{\rm sd}>10^{35}$~ergs~s$^{-1}$. 
For MSPs, the corresponding fraction is 8\%.
It is important to note, however, that these estimates are based on
observationally selected samples of radio pulsars which
are necessarily biased towards bright objects. As we show below, 
the true fraction of pulsars and MSPs in the population will be less.
In the following we estimate, using various arguments how many
pulsars and MSPs, may reside in the region of interest.

\subsection{The young NS population}

A simple argument can be made by noting that we have observed one
magnetar with projected distance of $\sim 0.1$~pc from Sgr~A* (Mori et
al.~2013; Eatough et al.~2013). In total there are 26 magnetars
currently known in the Milky Way (for a complete list, 
see Olausen \& Kaspi 2014
and the McGill Magnetar 
Catalog\footnote{http://www.physics.mcgill.ca/$\sim$pulsar/magnetar}).
Their typical lifetime from spin-down age arguments is $10^4$~yr, giving
a crude estimate on the magnetar birthrate to be one every $10^4/26
\sim 400$~yr. When compared to the pulsar birthrate of 2--3 per 
century (Vranesevic et al.~2004; Faucher-Giguere \& Kaspi 2006),
we conclude that for every 10 NS born, one of these will be a magnetar.
The existence of one magnetar in Sgr~A* therefore suggests the
existence of $\sim 10$~NS younger than $10^4$~yr.
Most of the pulsars in Fig.~4 with
characteristic ages below $10^4$~yr have 
$L_{\rm sd} > 10^{35}$~ergs~s$^{-1}$. Interestingly, those pulsars 
with ages below $10^4$~yr but $L_{\rm sd} < 10^{35}$~ergs~s$^{-1}$ 
are thought to be associated with the magnetar population:
J1550$-$5418 (Camilo et al.~2007),
J1622$-$4950 (Levin et al.~2010),
J1734$-$3333 (Espinoza et al.~2011).

An independent check on this simple estimate of $\sim 10$~NS with
$L_{\rm sd} > 10^{35}$~ergs~s$^{-1}$ can be made from 
the results of Chennamangalam \& Lorimer (2014) who found an
upper bound of 200 radio pulsars in the GC region that are beaming towards
the Earth. Assuming a beaming fraction of 10\% the total number of
radio-loud pulsars in this region is $\simless 2000$. Making use of
the PsrPopPy simulation software (Bates et al.~2014), we generated
a sample of radio pulsars which evolve with time according
to the prescription described in detail by Faucher-Giguere \& Kaspi (2006).
Under these assumptions, we find that the fraction of pulsars
with $L_{\rm sd}>10^{35}$~ergs~s$^{-1}$ is approximately 0.5\%.
The results of Chennamangalam \& Lorimer (2014) therefore limit
the population of $L_{\rm sd}>10^{35}$~ergs~s$^{-1}$ to be $<10$.
Relaxing the assumption of spin-down such that $L_{\rm sd}>10^{34}$~ergs~s$^{-1}$
would raise this limit by a further factor of three.
Given the uncertainties involved in both these estimates, we conclude
that one can reasonably expect of the order of several to a dozen
non-recycled pulsars of substantial spin down luminosity in the region
of interest.

\subsection{Millisecond pulsars}

In their population study, making use of the diffuse gamma-ray emission
observed by {\it Fermi}, Hooper \& Goodenough~(2011), 
Wharton et al.~(2012) estimated the
number of millisecond pulsars in the central 150~pc of Sgr~A* to
be $<5000$. Similar results were also found using earlier EGRET
observations by Wang et al.~(2005). To estimate the fraction of
millisecond pulsars potentially visible as X-ray transient sources,
we again used the PsrPopPy package to simulate a population
of pulsars spinning down from a spin period distribution
recently derived by Lorimer et al.~(2015) assuming a log-normal
distribution of magnetic field strengths with a mean of $10^8$~Gauss
and a standard deviation of 0.5 in the log. The fraction of 
sources with 
$L_{\rm sd}>10^{35}$~ergs~s$^{-1}$ is approximately 0.1\%, suggesting
a population of $<5$~sources. The corresponding fraction above
$10^{34}$~ergs~s$^{-1}$ is about 3\%, implying a population of
$<150$ millisecond pulsars.

\section{Discussion}

So far one magnetar is known within 0.1~pc from the GC.
There is also evidence that the source G359.95--0.04 
 at a projected distance of 0.3~pc from the GC is powered
by the wind of an energetic pulsar (Wang, Lu \& Gotthelf 2006). 
{We demonstrated above that $N\sim 10$  MSPs and 
pulsars with $L_{\rm sd} \simmore
10^{35}$~ergs~s$^{-1}$ may be expected to orbit arround Sgr A* 
within $\sim 0.1 \rm pc$. These pulsars spend most of their orbit at 
apocentre distance that determines their orbital period of 
$T\simeq 450 (r_{\rm app}/0.1\rm pc)^{3/2}$ years. Depending on the
ellipticity of their orbit, they can have brief excursions 
close to Sgr~A*. The stars observed at this scale -- the S cluster--
are characterized by highly elliptical orbits of median eccentricity
$e\simeq 0.8$ (Genzel et al.~2010). If the pulsars have similar 
orbits, their pericentre is $r_{\rm p} = (1-e)r_{\rm app}/(1+e)\simeq
0.01$~pc; well within the distance of interest here. 
The rate at which pulsars have their pericentre passage may be
 estimated to be  $\sim N/T\sim 2.2 N_1 (r_{\rm app}/0.1 {\rm pc})^{-3/2}$
per century.}

The relatively large gas density in this region, in connection with the large
 pulsar velocities results in strong interactions of the pulsar wind
 with the ambient gas. Pairs accelerated at the shocked pulsar wind
are strong synhrotron emitters. Given its high spatial resolution, {\it
Chandra} is well positioned to observe these interactions in the X-ray
band, {\it Fermi-LAT} may also be able to detect the source. 
Characteristic signatures include a flaring, non-thermal source
with luminosity $\simmore 10^{33}$~ergs~s$^{-1}$ that lasts for months or years.

{Other observed sources within the Chandra PSF at SGR A~* are the steady
 bremsstrahlung emission from $\sim 1$~keV gas and flares from the black-hole
vicinity (Baganoff et al. 2001). The flares last for minutes to hours
and are clearly distinct from the $\sim month-year$ long flares that
are discussed here. Month long flares from the same region may
originate from the interaction on {\it non}-relativistic winds of
massive stars with the disk (Giannios \& Sironi 2013). In this case, however, the 
flares are expected to be thermal and with their emission peaking in the soft x-ray band,
in contrast to the non-thermal transients discussed here.}

The interaction can serve as a probe of the density, temperature and thickness 
of the accretion disk that surrounds Sgr A~* at a scale  that is 
hard to probe otherwise (see also Nayakshin, Cuadra \& Sunyaev 2004;
Giannios \& Sironi 2013).
The duration of the event directly contrains the pericentre 
distance of the source $t_p\sim 3\times 10^7 R_{16}^{3/2}$~s.
Depending on the inclination of the NS orbit with respect to that
of the accretion disk, one or two flares are possible.
We encourage Chandra archival searches for such long transients
as well as systematic monitoring of Sgr~A*.

\section*{Acknowledgments}

This paper was initiated at a Scialog conference on Time Domain
Astrophysics held by the Research Corporation for Scientific 
Advancement. We thank the Research Corporation for their support of this work.

{}
\label{lastpage}


\begin{thebibliography}{}
\bibitem{a} Baganoff, F.~K, et al.\ 2001, Nature, 413, 45
\bibitem{b} Baganoff, F.~K., et al.\ 2003, ApJ, 591, 891
\bibitem{B} Bates, S.~D., et al.\ 2014, MNRAS, 439, 2893
\bibitem{c} Blandford, R.~D., \& Begelman, M.~C.\ 1999, MNRAS, 303, L1
\bibitem{d} Brandt, T.~D. \& Kocsis, B.\ 2015, ApJ, 812, 15
\bibitem{D} Camilo, F. et al.~2007, ApJ, 666, 93
\bibitem{e} Chennamangalam, J. \& Lorimer, D.~R.\ 2014, MNRAS, 440, L86
\bibitem{f} Cordes, J. \& Lazio, T.J.W.~1997, ApJ, 475, 557
\bibitem{g} Deneva, J.~et al.~2009, ApJ, 702, 17
\bibitem{h} de Jager, et al.\ 1996, ApJ, 457, 253
\bibitem{i} Dubus, G.\ 2013, A\&A Rev, 21, 64
\bibitem{j} Eatough, R.~et al.~2013, Nature, 501, 391
\bibitem{1} Espinoza C.~M.~et al.~2011, ApJ, 741, L13
\bibitem{J} Faucher-Giguere, C.A \& Kaspi, V.M.\ 2006, ApJ, 643, 332
\bibitem{k} Genzel, R., Eisenhauer, F. \& Gillessen, S.\ 2010, Rev. of Modern Physics, 82, 3121
\bibitem{l} Ghez, A.~M., et al.\ 2004, ApJ, 601, L159
\bibitem{m} Giannios, D. \& Sironi, S.~2013, MNRAS, 433, 25
\bibitem{n} Hooper, D. \& Goodenough, L.\ 2011, Physics Letters B, 697, 412
\bibitem{N} Levin, L. et al.\ 2010, ApJ, 721, L33
\bibitem{O} Lorimer, D.~R., et al.\ 2015, MNRAS, 450, 2185
\bibitem{o} Macquart, J-P. \& Kanekar, N.~2015, ApJ, 805, 172
\bibitem{p} Manchester, R.~N., et al.\ 2005, AJ, 129, 1993
\bibitem{q} Mori, K.~et al.~2013, ApJ, 770, 23
\bibitem{r} Nayakshin, S., Cuadra, J., \& Sunyaev, R.\ 2004, A\&A, 413, 173 
\bibitem{R} Olausen, S.~A. \& Kaspi, V.~M., ApJS, 212, 22
\bibitem{s} Paredes, J.~M., et al.\ 2013, Astroparticle Physics, 43, 301 
\bibitem{t} Pfhal, E. \& Loeb, A.~2004, ApJ, 615, 253
\bibitem{T} Porth, O., Komissarov, S.~S., \& Keppens, R.\ 2014, MNRAS, 438, 278 
\bibitem{u} Psaltis, D.~et al.~2015, ApJ, in press (arXiv:1510.00394)
\bibitem{v} Romani, R.~et al.~1997, ApJ, 484, 137
\bibitem{w} Tchekhovskoy, A. \& McKinney, J.~C.\ 2012, MNRAS, 423, L55
\bibitem{x} Qiang, Y. \& Zhang, B.~2014, JHEAp, 3, 1
\bibitem{y} Quataert, E. \& Gruzinov, A.\ 2000, ApJ, 539, 809 
\bibitem{z} van Eldick, C.~2015, APh, 71, 45
\bibitem{Z} Vranesevic, N. et al.\ 2004, ApJ, 617, L139
\bibitem{0} Wang, Q.~D., Lu, F.~J., \& Gotthelf, E.~V.\ 2006, MNRAS, 367, 937 
\bibitem{2} Wharton, R.~S., et al.\ 2012, ApJ, 753, 108
\end{thebibliography}
\end{document}